\documentstyle[prl,aps,twocolumn,epsf]{revtex}
\begin{document}

\input epsf
\draft
\twocolumn[\hsize\textwidth\columnwidth\hsize\csname
@twocolumnfalse\endcsname

\title{Signature of Quantum Hall Effect 
Skyrmions in Tunneling: A Theoretical Study} 
\author{J. J. Palacios and H.A. Fertig}
\address{Department of Physics and Astronomy, University of Kentucky,
Lexington, KY 40506, USA.}
\date{\today}
\maketitle

\widetext

\begin{abstract}

\leftskip 2cm
\rightskip 2cm

We present a theoretical study of the $I-V$ tunneling characteristic
between two parallel two-dimensional electron gases in a
perpendicular magnetic field when both are near filling factor $\nu=1$.
Finite-size calculations of the single-layer spectral functions in the
spherical geometry and analytical expressions for the disk
geometry 
in the  thermodynamic limit show that 
the current in the presence of skyrmions
reflects in a direct way their underlying structure. It is also shown
that fingerprints of the electron-electron interaction
pseudopotentials are present in such a current.

\end{abstract} 

\pacs{\leftskip 2cm PACS numbers: 73.23.Hk} 

\vskip2pc]

\narrowtext

Over the years, tunneling experiments have proven to be a powerful
source of information in many-body systems. The two-dimensional electron
gas (2DEG) in the quantum Hall regime\cite{girvin} is no exception.
Both edge and bulk tunneling densities of states have been probed
in different regimes of the bulk filling factor $\nu \equiv N/N_\phi$
($N$ is the number of particles and $N_\phi$ is the number of magnetic
flux quanta).  Recent tunneling measurements\cite{zhang} on edge states
have shown, for example, the first indications of chiral Luttinger
liquid behavior, which has been the focus of  much
theoretical\cite{edge:rev} and numerical\cite{jjp:lut} study.  For bulk
systems, a correlation pseudogap observed in tunneling
experiments\cite{exp:tun} has also motivated a large body of theoretical
work\cite{theory:tun}.

The region very near $\nu=1$, however, has received much attention
recently since the quasiparticles emerging from adding or removing
charges turn out to carry spin textures, and are commonly called
skyrmions\cite{lee,sondhi,fertig}.   In a simple characterization they
consist of a spin-polarized quasiparticle with $K$ (an integer)
additional electrons flipped into the minority
direction\cite{hcm,collmodes,jjp:sky,jacob}.  This in turn may be
interpreted as a charged quasiparticle with $K$ spin-waves bound to it.
The precise value of $K$ depends on the details of the electron-electron
interaction in the 2DEG as well as the strength of the Zeeman coupling.
It thus may be varied by tilting the sample with respect to the
direction of the magnetic field\cite{exp:sky}, or by more involved
methods\cite{maude}.  A strong experimental case that excess spins are
flipped when the quantum Hall ferromagnet is doped away from $\nu=1$ has
recently emerged\cite{exp:sky,maude}, with the corresponding value of
$K$ in remarkably good agreement with
theory\cite{fertig,jjp:sky,longhf}.  However, these experiments do not
actually show that the flipped spins are bound to the excess charges.
In this work, we show that tunneling experiments can directly
demonstrate this basic property of skyrmions, and measure, in a simple
way, both the binding energy and number of spin waves bound to the
quasiparticle.

We consider a system in which there are two parallel, weakly coupled,
2DEG's for which there are dilute and equal densities of quasiparticles
around the $\nu=1$ state.  For the low densities considered, a treatment
in which inter-skyrmion coupling\cite{brey} is neglected is appropriate.
Coulomb repulsions tend to keep the skyrmions in the same layer and
opposite layers apart, so that the problem reduces to considering
tunneling between a skyrmion and the $\nu=1$ ferromagnetic state.
Disorder effects are also ignored in our calculations. 

In addition to energy conservation, the conservation of spin greatly
limits the possible tunneling events that can occur.  For example, no
tunneling is possible between regions of pure $\nu=1$ ferromagnet
because no unoccupied states are available for the lowest Landau level
(LLL), spin majority electrons to tunnel into\cite{exp:tun}.  Tunneling
can {\it only} occur in the vicinity of the quasiparticles.  (In what
follows, we will consider for concreteness only quasihole excitations,
i.e., $\nu<1$.  The results for $\nu>1$ are identical due to an exact
particle-hole symmetry\cite{fertig,jjp:sky,longhf}.) For the case of
spin-polarized quasiparticles ($K=0$ skyrmions), this can only occur
precisely at zero voltage ($V=0$) between 2DEG's.  Thus, one expects a
$\delta$ function response in the $I-V$ curve which is presumably
broadened into a narrow peak by weak disorder.  

\begin{figure}
\hspace{2cm}
\epsfxsize=12cm \epsfbox{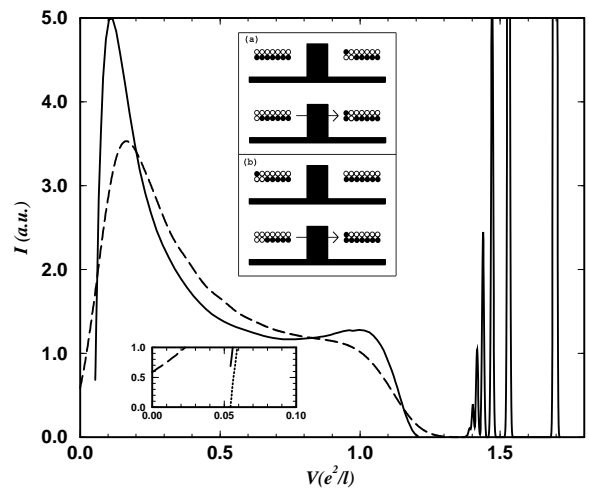}
\vspace{-4cm}
\caption{
Tunneling I-V in the presence of
skyrmions for disk (solid line)
and spherical (dashed line) geometries
(see text.)
Upper inset: Schematic representation  of the two possible tunneling
processes. (a) A majority-spin electron tunnels into a skyrmion.
(b) A minority-spin electron tunnels from the
skyrmion into a ferromagnetic region.  (Only the dominant
Slater determinant is shown.)
Lower inset: Blow-up of the low voltage region of $I-V$
using simple (solid line) and improved (dotted line) wavefunctions.
See text.}
\label{fig1}
\end{figure}

The case of $K=1$ quasiparticles, which is the major focus of this work,
is richer and far more interesting.  Fig.\ \ref{fig1} illustrates the
$I-V$ characteristic as calculated in several different ways,
and summarizes the central results of this work.  The dashed line
corresponds to an exact result obtained from a finite-size calculation
of electrons on a sphere.   A small broadening has been introduced into
the $\delta$-functions (which one always gets in any finite-size
calculation) to mimic the expected continuous curve of the thermodynamic
limit.  In sharp contrast to the $K=0$ case, the tunneling current has a
peak at {\it finite} $V$, and is suppressed near $V=0$.  However, in
order to understand the nature and functional form of this supression,
it is necessary to take the thermodynamic limit.
As will be shown below, this can be done {\it analytically}
in the disk geometry.   The result obtained from using a simple
wavefunction for the $K=1$ skyrmion\cite{jjp:sky} is shown as a
solid line in Fig.\ \ref{fig1}.  The tunneling current is completely
suppressed below a threshold voltage, and then jumps to a finite value.
This jump, however, turns out to arise from the fact that such a
wavefunction is not an eigenstate of $S^2$.  We propose below a new
$K=1$ wavefunction that {\it is} an eigenfunction of $S^2$.  The
resulting $I-V$ curve rises now linearly from the threshold voltage as
illustrated by the dotted line in the lower inset.

To understand the origin of the threshold, it is useful to consider the
forms of the initial and final states in the tunneling process [see
panel (a) in the upper inset]. A majority spin electron tunnels from the
ferromagnet and fills one of the holes of the $K=1$ skyrmion, liberating
one particle-hole pair.  Note that the hole left behind in the
ferromagnet is completely equivalent to a $K=0$ skyrmion state, i.e., a
spin-polarized quasiparticle.  The lowest energy state that can be a
final state on the right side is the $k \rightarrow 0$ spin-wave, which
has an interaction energy that is {\it identical} to the ferromagnetic
state.  Thus, the minimum energy that one must supply is the difference
in interaction energy between the $K=0$ and $K=1$ skyrmion states
$U_{K=0}-U_{K=1}$.  This is the threshold energy $eV_T$.  

We will argue below from symmetry considerations that for general values
of $K>0$ there will be a gap in the $I-V$, and that the onset of current
above the threshold takes the form $I(V) \propto [V-V_T]^{2K-1}$ for
$V>V_T$.  Here $V_T$ is the interaction energy gained by deforming a
spin-polarized quasiparticle into a skyrmion of size
$K$\cite{jjp:sky,longhf} ($U_{K=0}-U_K$).   Thus, tunneling
spectroscopy measures basic properties of the skyrmion in a direct way:
the voltage threshold for the onset of current is the binding energy of
the spin texture to the quasiparticle; the power law with which the
current first begins to flow measures the spin of the skyrmion.  For
experiments on non-disordered samples, this implies that the tunneling
$I-V$ characteristic changes sharply as the Zeeman coupling is varied
(as for example in tilted field geometries), whenever the equilibrium
value of $K$ changes.

Finally, the high-energy behavior of the tunneling $I-V$ curve displays
sharp peaks for $K=1$ skyrmions (Fig.\ \ref{fig1}).  These arise from
tunneling processes in which the minority electron tunnels out of the
skyrmion into the ferromagnet [panel (b) in upper inset].  The final
states on the skyrmion side are now polarized quasihole pairs, whose
eigenenergies are determined only by the relative angular momentum of
the two holes and are given by the pseudopotential parameters
$V_J$\cite{girvin} of the 2DEG.  It is worth noting that if these peaks
can be observed, this would constitute a direct measurement of $V_J$.
To our knowledge, there are no other such direct methods for measuring
these basic interaction parameters.

The tunneling current between the two parallel 
2DEG's is given to lowest order in
perturbation theory by 
\begin{equation}
I \propto \sum_{m,\sigma} \int_{-\infty}^\infty d\omega 
A^-_{m\sigma}(\omega-eV/2) A^+_{m\sigma}(\omega+eV/2),
\label{current}
\end{equation}
where the sum runs over all the single particle states in the LLL.  (Higher
Landau levels are not considered in this work.)
$A^{+(-)}_{m\sigma}$ are the single-layer
spectral functions for adding (removing) a
particle to (from) a state $m$ with spin $\sigma$.
In order to obtain the contribution to the current coming from the type
of processes depicted in Fig. 1, panel (a), we need four spectral
functions for each $m$: $A^+_{m\uparrow}, A^-_{m\uparrow},
A^+_{m\downarrow}$, and $A^-_{m\downarrow}$.  Taking $\uparrow$ as
majority spin, $ A^-_{m\downarrow}=0$, which leaves us only with the
spin-majority processes $A^+_{m\uparrow} A^-_{m\uparrow}$.  
For the ferromagnetic state and an appropriately chosen
zero of energy,
$A^-_{m\uparrow}(\omega) =\delta(\omega)$ and we only need to calculate
$A^+_{m\uparrow}(\omega)$.   To get the qualitative behavior of this
spectral function we first perform finite-size calculations on a  
sphere\cite{girvin}.  
For each value of the total spin operator $S^2$, single skyrmions are
obtained when $N=N_\phi-1$\cite{hcm,jjp:sky}. 
We will focus on the simplest one, $S=N/2-1$
($K=1$).  When such a skyrmion receives a spin-majority electron the
resulting state is a spin wave
whose form is known exactly in the spherical geometry\cite{aoki}, so
that calculation of $A^+_{m\uparrow}(\omega)$ is straightforward.  The
result of this calculation for $N=30$ corresponds to the dashed line
shown in Fig.\ \ref{fig1}. As expected, the major contribution comes
from injecting the electron at the center of the skyrmion which
gives rise to the peak around $V=0.1 e^2/\ell$.

Because $A^+_{m\uparrow}(\omega)$ is small for small $\omega$,
the behavior of the $I-V$ at low voltages
cannot be distinguished from finite-size 
studies.  In order to address this 
we turn our attention to the disk geometry.
The spin waves states of momentum $\vec k$ are given by
\begin{equation}
|\vec k\rangle=S^-(\vec k)|\Psi_0\rangle \equiv
{1 \over {\sqrt{N_{\phi}}}}\sum_X
e^{ik_xX}c^\dagger_{X+k_y\downarrow}c_{X\uparrow}|\Psi_0\rangle,
\label{spinwave}
\end{equation}
where the $c_{X\uparrow(\downarrow)}^{\dag}$'s create electrons in guiding
center coordinate states, and $|\Psi_0\rangle$ represents the
ferromagnetic state.  The magnetic length 
[$\ell \equiv \sqrt{\hbar c / (e B)}$] serves as our unit of length.
The corresponding energies, $\epsilon(k)$, are also easily 
evaluated\cite{kallin}.
We need to compute
the overlap of these states with $c_{m\uparrow}^{\dag}|K=1\rangle$,
where $|K=1\rangle$ represents the skyrmion state.
In general, the neutral object $c_{m\uparrow}^{\dag}|K=1\rangle$
may always be written
in the form 
$\sum_{m_e,m_h} \alpha^{m_\uparrow}_{m_e m_h} |m_\uparrow;m_e m_h \rangle$,
i.e., as a sum of states with one hole in the majority spin ($m_h$)
and one electron  ($m_e$) in the minority spin.
Written in this form, the matrix element 
$\langle\vec k|c^{\dag}_{m\uparrow}|K=1\rangle$ may be computed exactly,
and, furthermore, the sum over all final spin wave states
in the spectral function can be computed analytically in the
thermodynamic limit.  The result of this calculation takes the form
\begin{eqnarray}
A^+_{m\uparrow}(\omega)&=&k_0e^{-k_0^2/2}\left|\frac{1}{d\epsilon(k)/dk}
\right|_{k=k_0} \nonumber \\
&&\times \sum_{m_e,m_h,m'_e,m'_h} \alpha^{m_\uparrow}_{m_e m_h}
(\alpha^{m_\uparrow}_{m'_e m'_h})^* \nonumber \\
&& \times \delta_{m_e-m_h,m'_e-m'_h}
(i)^{m_e+m_h}(-i)^{m'_e+m'_h} \nonumber \\
&& \times k_0^{m_e+m_h+m'_e+m'_h} 
\left[\sum_{n=0}^M (-1)^n \beta^n_{m_e m_h} k_0^{-2n}\right] \nonumber \\
&&\left[\sum_{n'=0}^{M'} (-1)^{n'} \beta^{n'}_{m'_e m'_h} k_0^{-2n'}\right],
\label{specfun}
\end{eqnarray}
where $k_0$ is
implicitly given by $\epsilon(k_0)=\omega$, $\beta_{m_1,m_2}^n=
\left(\begin{array}[c]{c}m_1\\n\end{array}\right)
\left(\begin{array}[c]{c}m_2\\n\end{array}\right) 2^n
n!/\sqrt{2^{m_1+m_2}m_1!m_2!}$,
and $M$ is the largest integer for which $\beta_{m_1,m_2}^M$
is defined.

In order to use Eq. \ \ref{specfun}, one needs to have
an explicit form for the skyrmion wavefunction, which supplies
the coefficients $\lbrace\alpha\rbrace$.
For the $K=1$ skyrmion, the simplest
variational wave function that has $K$ as a good quantum number 
looks schematically like\cite{jjp:sky}
\begin{eqnarray}
|K=1\rangle_{\gamma}&\equiv&
\gamma_1 
\left| \left. 
\frac{\bullet\circ\circ\circ\circ\circ\dots}
{\circ\circ\bullet\bullet\bullet\bullet\dots}
\right. \right)
+
\gamma_2
\left| \left. \frac{\circ\bullet\circ\circ\circ\circ\dots}
{\circ\bullet\circ\bullet\bullet\bullet\dots}  \right. \right)
+ \nonumber \\
&&\gamma_3
\left| \left. \frac{\circ\circ\bullet\circ\circ\circ\dots}
{\circ\bullet\bullet\circ\bullet\bullet\dots}  \right. \right)
+ \dots,
\label{sky:alpha}
\end{eqnarray}
where the upper (lower) circles represent the spin-minority (-majority) 
single-particle states in the circular gauge,
and the optimal coefficients have been found to be given by
$\gamma_j \propto \frac{e^{-0.071j}}{\sqrt{j}}$\cite{jjp:sky}. 

The values of $\alpha^{m_\uparrow}_{m_e m_h}$
can be obtained straightforwardly for this wavefunction, and the resulting
$I-V$ is presented in Fig.\ \ref{fig1} (solid line). 
It is interesting to note
that in this approximation we find 
$A^+_{m\uparrow}(\omega) \propto \omega^{|m-1|}$, so that
all the spectral functions vanish as $\omega \rightarrow 0$
except for the case $m=1$, leading to the jump in current
at $V_T$.  However, a symmetry analysis shows that
this behavior cannot be correct: all the
spectral functions $A^+_{m\uparrow}(\omega)$ must
vanish as $\omega \rightarrow 0$.  To see this, consider
the matrix element $\langle \vec k|c^{\dag}_{m\uparrow}|K=1\rangle$.
The spin wave operator $S^-(\vec k)$ (Eq.\ \ref{spinwave})
commutes with $c^{\dag}_{m\uparrow}$ so that this expression
may formally be rewritten as $\langle K=0|S^-(\vec k)|K=1\rangle$ ($|K=0
\rangle$ represents a polarized quasihole state). 
In the limit $\omega \rightarrow 0$ only the $\vec k \rightarrow 0$
matrix element enters $A^+_{m\uparrow}(\omega)$.
However, $S^-(0)$ is just the spin lowering operator, which
can change only the $S_z$ eigenvalue of any eigenstate
of the Hamiltonian, but not its total spin $S$ eigenvalue.
Since skyrmions of different $K$ values are in different
$S^2$ multiplets\cite{hcm}, it follows that 
$\langle K=0|S^-(\vec k)|K=1\rangle$ vanishes as $k \rightarrow 0$,
presumably as $k$, the lowest order allowed by symmetry.
Noting that the spin wave spectrum of the ferromagnet
has energy state $\propto k^2$ for small $k$, one can
easily see from Eq.\ \ref{specfun} that all the spectral
functions must vanish at least linearly with $\omega$
for the $K=1$ skyrmion.

That Eq.\ \ref{sky:alpha} above does not respect this
property is due to the fact that it is not an eigenfunction
of $S^2$.  We propose an improved wavefunction that is
in the appropriate $S^2$ multiplet:
\begin{eqnarray}
|K=1\rangle_{\gamma,\gamma'}& \equiv&
\gamma_1 
\left| \left. 
\frac{\bullet\circ\circ\circ\circ\circ\dots}
{\circ\circ\bullet\bullet\bullet\bullet\dots}
\right. \right)
+
\gamma_2
\left| \left. \frac{\circ\bullet\circ\circ\circ\circ\dots}
{\circ\bullet\circ\bullet\bullet\bullet\dots}  \right. \right)
+ \nonumber \\
&&\gamma_3
\left| \left. \frac{\circ\circ\bullet\circ\circ\circ\dots}
{\circ\bullet\bullet\circ\bullet\bullet\dots}  \right. \right)
+ \dots
+ \nonumber \\
&&\gamma'_1 
\left| \left. 
\frac{\circ\circ\bullet\circ\circ\circ\dots}
{\bullet\circ\circ\bullet\bullet\bullet\dots}
\right. \right)
+
\gamma'_2
\left| \left. 
\frac{\circ\circ\circ\bullet\circ\circ\dots}
{\bullet\circ\bullet\circ\bullet\bullet\dots}  
\right. \right)
+ \nonumber \\
&&\gamma'_3
\left| \left. 
\frac{\circ\circ\circ\circ\bullet\circ\dots}
{\bullet\circ\bullet\bullet\circ\bullet\dots} 
 \right. \right)
+ \dots,
\label{sky:alphagamma}
\end{eqnarray}
with the condition $\gamma_1+\sum_{m=1}^\infty\gamma'_m=0$
guaranteeing that Eq.\ \ref{sky:alphagamma} is in a well-defined
$S^2$ multiplet.  The coefficients $\lbrace \gamma_m,\gamma'_m \rbrace$
can be  obtained numerically by diagonalizing
the Hamiltonian in such a Hilbert subspace.
Table I illustrates the overlaps of the two approximate wavefunctions
with exact-diagonalization wavefunctions on a finite disk.  Although both
states have very large overlaps, the $|K=1\rangle_{\gamma,\gamma'}$
wavefunctions are clearly two orders of magnitude closer to
the exact wavefunctions.  
The dotted line  in lower inset shows the I-V curve at the
onset of current for
$|K=1\rangle_{\gamma,\gamma'}$
with the coefficients extrapolated to
the thermodymanic limit.   The current vanishes linearly near $V_T$
since now $A^+_{1\uparrow}$ has the proper behavior; the 
rest of the curve is essentially identical to the result obtained
using the $|K=1\rangle_{\gamma}$ wavefunction.

The contribution to the I-V characteristic
coming from the spin-minority processes shown in panel (b), upper inset,
dominates the voltage region above $\sqrt{\pi/2}e^2/\ell$. When a
minority-spin electron is removed from a $K=1$ skyrmion two
quasiholes are left behind. 
The eigenstates and energies of
this system  
have well-defined relative ($J$=odd integer) and center of mass
angular momentum; the energy of the state depends
only on $J$ and is given by the Haldane
pseudopotential parameters $V_J$\cite{girvin}.  
The
weights associated with these states are easily calculated;
for the $|K=1\rangle_{\gamma}$ one finds
$$
W_J=\sum_{m=J-1}^\infty 2^{-m} \gamma_{m+1}^2
\frac{(m+1)!}{J!(m+1-J)!}. 
$$
The results are shown in Fig.\ \ref{fig1} where the delta functions have
been broadened to make them visible.  

What about skyrmions with larger values of $K$?  Exact analytic
expressions for the spectral functions cannot be obtained using
the methods above.  However, we can draw some conclusions about
what they and the resulting $I-V$ must look like. 
Based on the previous reasoning, there
will be a gap in the tunneling $I-V$, with current
starting to flow at $eV_T=U_{K=0}-U_{K}$.
For voltages just above $V_T$, the final state on the skyrmion
side may be interpreted as a linear combination of $K$ spin waves,
all with small wavevectors.  Such states may be written in
the form $\prod_{i=1}^{K} S^-(\vec{k}_i) |\Psi_0\rangle$.  Noting
again that $c_{m\uparrow}^\dagger$ commutes with $S^-(\vec{k}_i)$,
the matrix elements entering the spectral functions may be written
as $\langle K=0|\prod_{i=1}^{K} S^-(\vec{k}_i) |K\rangle$.  {\it All} the 
$\vec{k}_i$'s must be non-zero in order for this expression
to be non-vanishing, just as in the case described above
for $K=1$.  Thus, the matrix elements entering into Eq. \ref{current}
should be proportional to $\prod_i k_i^2$.  Assuming that the
spin waves may be treated as non-interacting in the
long-wavelength limit, the energy of the $K$ spin wave state
will be $\propto \sum_i k_i^2$.  Using these observations, 
one finds that $A^+_{m\uparrow}(\omega)$
must vanish at least as fast as $(\omega-V_T)^{2K-1}$.  The resulting
current then rises from $V_T$ as $(V-V_T)^{2K-1}$.  
Thus, for any skyrmion with $K>1$,
the threshold voltage and power law for the onset of current are respectively
measures of the binding energy of the spin texture
and the total number of flipped spins bound to it.


\noindent {\it Acknowledgements --} It is a pleasure to thank A. H.
MacDonald for helpful comments.  This work has been supported by 
NSF Grant DMR-9503814 and the Research Corporation.

\begin{table}
\caption{Overlaps between the exact
skyrmions and the trial states $|K=1\rangle_{\gamma}$ and 
$|K=1\rangle_{\gamma,\gamma'}$.}

\begin{tabular}{|c|ccc|}

& $N=5$  & $N=10$ & $N=15$ \\ \hline
$|K=1\rangle_{\gamma}$ & 0.940960  & 0.974519 & 0.983732 \\
$|K=1\rangle_{\gamma,\gamma'}$ & 0.999630 &  0.999703 & 0.999658 
\end{tabular}
\end{table}

\vspace{-5.5cm}

\end{document}